\documentclass[journal]{IEEEtran}
\usepackage{amsmath,amsfonts}
\usepackage{algorithmic}
\usepackage{algorithm}
\usepackage{array}
\usepackage[caption=false,font=normalsize,labelfont=sf,textfont=sf]{subfig}
\usepackage{textcomp}
\usepackage{stfloats}
\usepackage{url}
\usepackage{verbatim}
\usepackage{graphicx}
\usepackage{cite}
\usepackage{hyperref}
\usepackage{xspace}
\usepackage{booktabs}
\usepackage{tabu}
\usepackage{mathptmx}
\usepackage{wrapfig}
\usepackage{scalerel}
\usepackage[normalem]{ulem}
\usepackage{soul}
\usepackage[dvipsnames,svgnames,x11names]{xcolor}
\usepackage{ifthen}
\usepackage{etoolbox}
\usepackage{enumitem}  
\usepackage[capitalise,noabbrev]{cleveref}  

\newcommand{\eg}{\textit{e.g.},\xspace}
\newcommand*{\icon}[1]{\scalerel*{\includegraphics{figs/icons/#1}}{\strut}}
\providecommand{\paperVersion}{arxiv}
\newif\ifarxivversion
\newif\ifreviewversion
\newif\ifjournalversion
\ifthenelse{\equal{\paperVersion}{arxiv}}{\arxivversiontrue}{%
        \ifthenelse{\equal{\paperVersion}{review}}{\reviewversiontrue}{%
                \ifthenelse{\equal{\paperVersion}{journal}}{\journalversiontrue}{%
                        \PackageError{vispilot}{Unknown paperVersion `\paperVersion'}{Use arxiv, review, or journal.}%
                }%
        }%
}

\ifarxivversion
\patchcmd{\IEEEbiographynophoto}
  {\vskip 4\baselineskip plus 1fil minus 0\baselineskip}
  {\vskip 2\baselineskip plus 1fil minus 0\baselineskip}
  {}{\PackageError{vispilot}{Unable to adjust biography spacing}{Check the IEEEtran class version.}}
\fi

\newcommand{\bodycolor}[1]{\ifarxivversion\color{#1}\else\color{Black}\fi}
\newcommand{\revisioncolor}[1]{\ifreviewversion\color{#1}\else\color{Black}\fi}

\newcommand{\dataprop}[1]{{\bodycolor{CornflowerBlue}#1}}
\newcommand{\markprop}[1]{{\bodycolor{Coral}#1}}
\newcommand{\encprop}[1]{{\bodycolor{Teal}#1}}
\newcommand{\designprop}[1]{{\bodycolor{MediumPurple}#1}}

\newcommand{\rev}[1]{{\color{Black}#1}}
\newcommand{\revadd}[1]{{\revisioncolor{BrickRed}#1}}

\newcommand{\fref}[2]{\hyperref[{#1}]{\cref*{{#1}}-{#2}}}
\newcommand{\hfref}[2]{\hyperref[{#1}]{{#2}}}

\graphicspath{{figs/}{figures/}{pictures/}{images/}{./}}

\ifdefined\pdfminorversion\pdfminorversion=7\fi
\hypersetup{
    pdftitle={Exploring Multimodal Prompt for Visualization Authoring with Large Language Models},
    pdfauthor={Zhen Wen, Luoxuan Weng, Yinghao Tang, Runjin Zhang, Yuxin Liu, Bo Pan, Minfeng Zhu, Wei Chen},
    pdfsubject={Author-prepared version for arXiv},
    pdfkeywords={visualization authoring, large language models, multimodal prompting},
    hidelinks
}

\begin{document}

\title{Exploring Multimodal Prompt for Visualization Authoring with Large Language Models}

\author{Zhen~Wen,
        Luoxuan~Weng,
        Yinghao~Tang,
        Runjin~Zhang,
        Yuxin~Liu,
        Bo~Pan,
        Minfeng~Zhu,
        and~Wei~Chen%
\thanks{
This work is supported by the National Natural Science Foundation of China under Grant No. 62421003 and No. 62302435, and in part by the Zhejiang Provincial Natural Science Foundation of China under Grant No. LD24F020011.

Zhen Wen, Luoxuan Weng, Yinghao Tang, Runjin Zhang, Yuxin Liu, Bo Pan, and Wei Chen are with the State Key Lab of CAD\&CG, Zhejiang University. E-mail: \{wenzhen, lukeweng, yinghaotang, 3210105680, 3220104160, bopan, chenvis\}@zju.edu.cn.}%
\thanks{Minfeng Zhu is with Zhejiang University. E-mail: minfeng\_zhu@zju.edu.cn.}
\thanks{Zhen Wen and Luoxuan Weng contributed equally to this work.}
\thanks{Corresponding authors: Minfeng Zhu and Wei Chen.}%
\ifarxivversion
\thanks{\raggedright \textcopyright\ 2026 IEEE. Personal use permitted; republication/redistribution requires \href{https://www.ieee.org/publications/rights/index.html}{IEEE permission}.}%
\fi}

\markboth{Journal of \LaTeX\ Class Files,~Vol.~14, No.~8, August~2021}%
{Shell \MakeLowercase{\textit{et al.}}: A Sample Article Using IEEEtran.cls for IEEE Journals}

\IEEEpubid{}

\maketitle

\begin{abstract}
Recent advances in large language models (LLMs) have shown great potential in automating the process of visualization authoring through simple natural language utterances. However, instructing LLMs using natural language is limited in precision and expressiveness for conveying visualization intent, leading to misinterpretation and time-consuming iterations. To address these limitations, we conduct an empirical study to understand how LLMs interpret ambiguous or incomplete text prompts in the context of visualization authoring, and the conditions making LLMs misinterpret user intent. Informed by the findings, we introduce visual prompts as a complementary input modality to text prompts, which help clarify user intent and improve LLMs' interpretation abilities. To explore the potential of multimodal prompting in visualization authoring, we design VisPilot, which enables users to easily create visualizations using multimodal prompts, including text, sketches, and direct manipulations on existing visualizations.
\revadd{We evaluate VisPilot through a controlled user study and an expert evaluation. 
The results suggest that multimodal prompts facilitate users in communicating spatial constraints, local references, and design preferences while maintaining comparable task efficiency to text-only prompting. 
We further discuss when text, visual, and hybrid prompts are beneficial for visualization authoring, and summarize design implications for future human-AI authoring systems.} 
All materials are available at \url{https://osf.io/2qrak}.
\end{abstract}

\begin{IEEEkeywords}
Visualization authoring, large language model, multimodal prompting.
\end{IEEEkeywords}

\section{Introduction}
\IEEEPARstart{V}{isualization} authoring tools have evolved rapidly to lower the barriers of creating data visualizations, from expertise-driven languages and formal grammars~\cite{Bostock2011D3,Satyanarayan2017VegaLite,LYi2022Gosling} to more accessible graphical interfaces~\cite{Wongsuphasawat2016Voyager,Wongsuphasawat2017Voyager2,Moritz2019Draco,Wang2024AVA,Wang2024DataFormulator,Han2024Nuwa}. 
With the emergence of large language models (LLMs), visualization creation has been further simplified through natural language interfaces that automatically translate user utterances into visualization specifications~\cite{dibia2023lida,Ye2024Generative,han2023chartllama,tian2024chartgpt}. 
However, despite their accessibility, recent studies indicate that LLMs are limited in understanding accurate visual intent from natural language inputs~\cite{shen2025prompting,chen2025interchat}.

Conveying visualization intent in natural languages faces challenges in terms of accuracy and expressiveness~\cite{Yi2025Blace}. 
First, natural language inherently contains {\it ambiguity and implicit cues}~\cite{Srinivasan2021Snowy,Feng2024XNLI,luo2025nvbench}, requiring LLMs to infer the user's true intent. 
Since such inferences can be ambiguous, the visualizations often deviate from the expected outcome.
Second, there exists a fundamental {\it modality gap} between textual descriptions and graphical visualizations: users struggle to precisely articulate visual intent through text alone.
Once the model generates an unexpected visualization, it is difficult for users to diagnose or correct the underlying issues through text-only instructions. 
\revadd{This gap becomes especially salient in iterative authoring, where users often need to refer to a local element in the current chart, specify a spatial relationship, draw a boundary for filtering, or demonstrate a visual style by example. Such intents are easy to express by pointing, circling, sketching, or annotating in place, but they require verbose and fragile descriptions in a conversational text interface.} 
These limitations highlight the need for a more effective approach to instruct LLMs with accurate visual intent in the process of visualization authoring. 

To enable LLMs to more accurately understand visual intent, recent research has explored multimodal prompting in the tasks of image generation~\cite{DBLP:conf/nips/NguyenLOL23} and visual question answering~\cite{DBLP:conf/eccv/WangG24}.
Multimodal prompting combines textual and visual inputs to enhance the understanding of user intent, allowing for more accurate and expressive outputs.
\revadd{For visualization authoring, however, the challenge is not only to send an image to a multimodal model, but also to ground visual cues in data attributes, chart elements, and specification-level constraints.}
\revadd{It remains unclear when visual prompts add value over natural language, and how an authoring interface should guide users so that free-form sketches become interpretable visualization instructions rather than ambiguous marks on a canvas.}
\rev{While some diffusion-based work have utilized multimodal inputs for generating content~\cite{Zhang2023ControlNet,Schetinger2023Doom}, visualization requires a high degree of structural rigour and data fidelity that pixel-based generation cannot provide.}
\rev{This lack of precision is particularly problematic in professional data analysis and journalism, where users often have specific mental models of layouts and encoding designs that are difficult to verbalize.} 

\begin{figure*}[t]
\centering
\includegraphics[width=\textwidth]{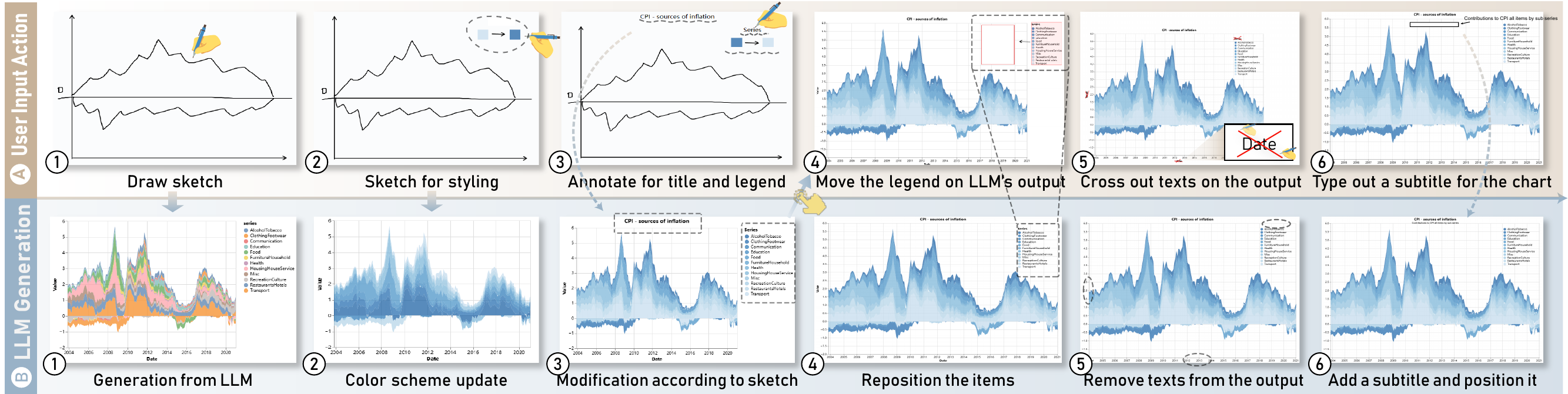}
\caption{Multimodal prompt for visualization authoring with VisPilot. (A) The user can create visualizations by providing sketching, text annotations or directly manipulating existing visualizations. (B) VisPilot interprets the multimodal input and generates visualizations.}
\label{fig:teaser}
\end{figure*}

We conduct an empirical study to understand the limitations of using text prompts to instruct LLMs for visualization creation.
Through a systematic analysis of 814 natural language utterances used to request visualizations, we find that prompting LLMs with natural language frequently leads to misinterpretation of user intent due to three reasons:
(1) limited expression of visual intent: text prompts are inherently inflexible to express visual intent;
(2) inadequate guidance for LLM behavior: non-expert or inaccurate expressions constantly lead to LLMs' incorrect inferences;
(3) misaligned human-LLM design preferences: prompts without explicit design instructions result in biased design preferences of different LLMs.
\revadd{These findings motivate a hypothesis on whether visual prompts can provide complementary cues for intents that are difficult to verbalize, such as spatial constraints, references to local chart elements, and visual designs.}
\revadd{Building on this motivation,}
we propose a multimodal prompting framework that incorporates visual prompts as a complement to text prompts.

We develop VisPilot as a proof-of-concept system to evaluate the feasibility of the multimodal prompting framework. 
The system allows users to create visual prompts by sketching on a canvas. 
VisPilot supports four fundamental visual input actions: scratch (freehand drawing of visualization elements), style (applying visual properties), annotation (providing textual context), and manipulation (modifying existing elements). 
This suite of interactions enables users to express visualization intent through visual manners, complementing traditional text prompts. 
\revadd{Unlike existing visual analytics systems~\cite{Masson2024DirectGPT,chen2025interchat,Wang2024DataFormulator} with multimodal interaction that map interactions to fixed widgets or hidden text templates, and unlike free-form sketch-only approaches~\cite{Yen2025CodeShaping} that leave the interpretation of drawings largely to the model, our framework treats visual prompts as first-class reasoning cues while grounding them with visualization-specific operations and interaction descriptions.}
The underlying prompting framework interprets these visual inputs through a step-wise reasoning process, translating sketches into precise visualization specifications.
We evaluate VisPilot through a controlled user study comparing text-only and multimodal prompting for visualization authoring tasks.
The results indicate that VisPilot can reduce errors and improve user satisfaction without significantly affecting task efficiency.
\revadd{The expert evaluation further deepens the analysis by showing how professional users employ visual prompts for spatial constraints, local references, and visual designs in open-ended authoring workflows.}
Based on the findings from evaluation studies, we identify several valuable design implications for multimodal prompting in visualization authoring scenarios.

In summary, our main contributions include:
\revadd{(1) a formative empirical study identifying how text-only prompts lead LLMs to make implicit and often erroneous inferences for visualization specifications,}
\revadd{(2) a guided visual prompting framework and prototype system}\footnote{An online demo is available at \href{https://wenzhen.site/vispilot}{https://wenzhen.site/vispilot}.}\revadd{ that operationalize visual prompts as structured, visualization-specific reasoning cues,}
\revadd{and (3) an evaluation analyzing where text, visual, and hybrid prompts help users specify visualizations, along with design implications for future multimodal visualization authoring systems.}

\section{Related Work}

\subsection{LLMs for Visualization Generation}

LLMs have been demonstrated as a convenient interface for natural language to visualization tasks~\cite{Wu2024AutoVis,Li2024VisGenLLM,Wang2023VisAI}. 
Recent research has focused on enhancing LLMs for visualization generation, with a particular emphasis on model fine-tuning and prompting strategies.
ChartLlama~\cite{han2023chartllama} and ChartGPT~\cite{tian2024chartgpt} fine-tune language models with visualization domain knowledge, while systems like LIDA~\cite{dibia2023lida} and FinFlier~\cite{Hao2024FinFlier} leverage carefully designed prompting strategies for visualization generation without modifying the underlying models.
Several studies have proposed evaluation criteria for LLM-generated visualizations. VisEval~\cite{Chen2025VisEval} and DracoGPT~\cite{Wang2025DracoGPT} establish benchmarks and metrics to assess visualization quality, appropriateness, and adherence to design principles.
\revadd{Recent surveys on generative AI for visualization~\cite{Ye2024Generative} and visualization-supported multimodal analysis~\cite{Wang2025MultimodalAnalysisSurvey} further show that visualization research is increasingly coupled with multimodal data, models, and interaction workflows.}

While these approaches have advanced LLM-based visualization generation, they primarily formulate the visualization requirement as a natural language prompt, which may not fully capture the complexity of user intent and design principles.
Our work aims to address this gap by exploring multimodal prompting strategies that incorporate both visual and textual elements to enhance the generation of visualizations.

\subsection{Multimodal Prompt for Generative Models}

Due to the inherent ambiguity and redundancy of text prompts, researchers have explored multimodal prompt design for LLMs~\cite{DBLP:journals/tvcg/ZengLYZ25}.
For example, visual prompts like colorful boxes or circles can direct multimodal large language models (MLLMs) to specific regions of interest, thereby improving their generation quality~\cite{Wu2024ControlMLLM, DBLP:conf/nips/YangWLWY23, DBLP:journals/corr/abs-2312-05278}. This strategy has been widely applied to computer vision tasks such as visual question answering~\cite{DBLP:conf/eccv/WangG24}, image editing~\cite{DBLP:conf/nips/NguyenLOL23}, and knowledge tagging~\cite{DBLP:conf/cvpr/GuptaK23}.
Recent studies have also investigated interaction-augmented prompts to facilitate precise user intent understanding~\cite{shen2025prompting}. Chen \emph{et al.}~\cite{chen2025interchat} propose a design space for generative visual analytics and develop a direct manipulation interface. Similarly, DirectGPT~\cite{Masson2024DirectGPT} characterizes four direct manipulation actions to enhance the efficiency of human-LLM communication. However, these approaches typically transform interactions back to engineered text prompts, without explicitly incorporating visual inputs. Moreover, the current literature lacks a clear understanding of multimodal prompt design considerations to improve the expressiveness and efficiency of the visualization authoring process. 
\revadd{Code Shaping~\cite{Yen2025CodeShaping} demonstrates that free-form sketches can guide AI-assisted code editing, providing an important baseline for sketch-based instruction. 
In visualization authoring, however, sketches need to be grounded in data attributes, chart elements, and specification-level constraints, calling for structured, visualization-specific cues beyond free-form marks alone.}
Our work extends prior endeavors by identifying key design principles for visual prompting and examining how different prompt modalities influence visualization specification.

\subsection{Multimodal Interactions for Visualization}

Multimodal interactions have been widely studied in the context of visualization systems, enabling users to create or interact with visualizations through multiple input modalities such as direct manipulations~\cite{Saket2018Evaluating}, freehand sketches~\cite{Srinivasan2021Interweaving}, and natural language queries~\cite{Saktheeswaran2020Touch}.
Due to the accessibility and extensibility of natural language, many systems have explored combining verbal and visual modalities to enhance user experience in visualization authoring. DataTone~\cite{Gao2015DataTone} and Orko~\cite{Srinivasan2018Orko} pioneered approaches that integrate natural language with direct manipulation interfaces, allowing users to refine ambiguous queries through interactive widgets. Similarly, tools like Valletto~\cite{Saktheeswaran2020Touch,Kassel2018Valletto} facilitate natural language interactions with visualizations through contextual dialogs and touch-based interactions.
Recent systems such as WYTIWYR~\cite{Xiao2023WYTIWYR} recommend visualizations based on user intent and multimodal inputs, while VisTR~\cite{Hao2024VisLTR} aligns charts, text, and sketches with visualization representations to support time-series table reasoning.
These studies have demonstrated the potential of multimodal interactions in enhancing user experience on data visualization and exploration.
\revadd{Existing surveys also emphasize that visualization is valuable for integrating and reasoning across heterogeneous modalities~\cite{Wang2025MultimodalAnalysisSurvey}.}
Informed by them, we aim to investigate how multimodal prompting can be effectively utilized in LLMs for visualization authoring.

\section{Empirical Study: Text Prompt for Visualization}

To understand the limitations of text prompts for visualization authoring, we conduct a systematic analysis of 814 natural language utterances used to request visualizations.
The analysis identifies the key challenges that multimodal prompting approach needs to address.
In particular, our empirical study aims to address the following research questions:
\begin{itemize}[leftmargin=*]
    \item
    \emph{\textbf{RQ1:} How do LLMs interpret natural languages for visualization specifications?}
    The natural language utterance for creating a visualization often expresses multifaceted preferences for visualization design, such as mark type, encoding choices, etc.
    Previous studies~\cite{Arjun2021Collecting,Li2024VisGenLLM} have shown that users often explicate only partial specifications in their utterances.
    These observations raise the question of how LLMs infer complete visualization specifications from ambiguous or incomplete natural language utterances.
    \item 
    \emph{\textbf{RQ2:} What makes LLMs generate unexpected visualization specifications?}
    The performance of LLMs in generating visualization specifications is often unpredictable.
    While recent studies~\cite{Chen2025VisEval,Wu2024AutoVis} have discussed the reasons behind unexpected results in terms of prompt design and model capabilities, the effects of ambiguity and incompleteness of user utterances are still unclear.
    We aim to understand how the ambiguity and incompleteness in natural language expressions contribute to unexpected visualization outputs.
\end{itemize}

\begin{figure}[b]
    \centering
    \includegraphics[width=0.99\columnwidth]{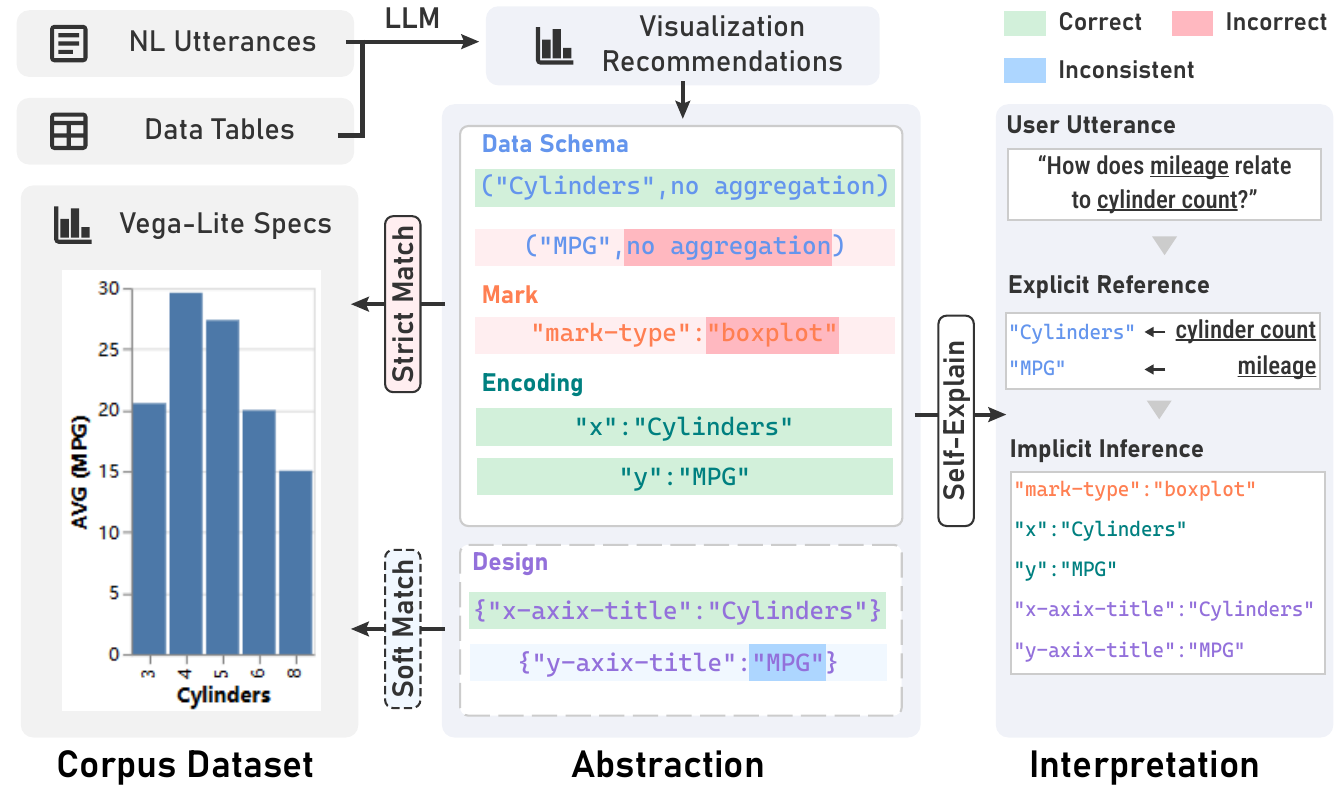}
    \caption{The procedure of LLM processing on the corpus data. \rev{The LLM generates visualizations from natural language utterances, which are abstracted into four specification components (Data Schema, Mark, Encoding, Design). These components are compared against ground truth using strict or soft constraints to determine accuracy, while the LLM self-explains its generation rationale (from explicit references or implicit inferences).}}
    \label{fig:corpus-pipeline}
\end{figure}

\begin{figure*}[t]
    \centering
    \includegraphics[width=0.99\linewidth]{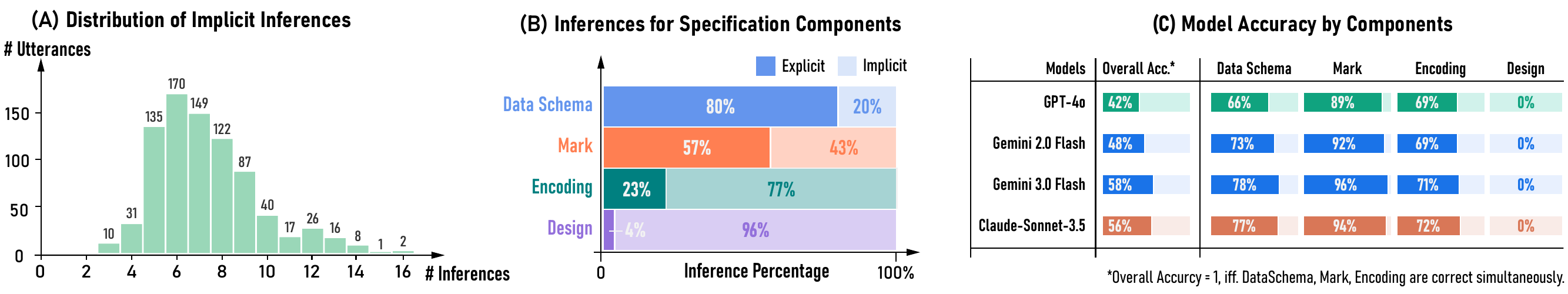}
    \caption{The analysis results of LLM interpretation of utterances. (A) The number of implicit inferences made for each utterance, where each item of specification components is counted separately.
    (B) The percentage of explicit references or implicit inferences made for each specification component, where a component is counted as implicit if at least one of its items is implicitly inferred.
    \rev{(C) The accuracy of generated specifications across four models.}
    }
    \label{fig:nl-queries}
\end{figure*}

\subsection{Methodology}

\noindent
\textbf{Data Source.}
We analyze the corpus of 814 natural language utterances collected by Srinivasan et al.~\cite{Arjun2021Collecting}. 
This dataset was compiled through a structured study with 102 participants who were shown ten canonical visualizations (\eg bar charts, line charts, scatterplots) and asked to provide natural language utterances they would use to create these visualizations. 
Each utterance in this corpus is paired with both the data source and the target visualization written in Vega-Lite code.

\vspace{2pt}
\noindent
\textbf{Framework.}
To better understand how LLMs interpret natural language utterances for multifaceted visualization specifications, we establish a framework to abstract the components of specifications referring to previous work~\cite{Moritz2019Draco,Satyanarayan2017VegaLite}.
This framework serves as a foundation for our analysis of the corpus and the design of multimodal prompting approach.
We define a specification $S$ as a composition of four basic components that together determine the complete visualization design:

\begin{itemize}[leftmargin=*]
    \item[{\bodycolor{CornflowerBlue}\small$\bullet$}]
    \textbf{\bodycolor{CornflowerBlue}Data Schema $\mathcal{D}$.} 
    This component defines the structural properties of data, including attribute names and data transformations. 
    Formally, $\mathcal{D} = \{(a_i, f_i)\}$ where $a_i$ is an attribute, and $f_i$ is an optional aggregation function (\eg sum, average, count). 
    The schema provides a foundation for understanding the data's structure and semantics.

    \item[{\bodycolor{Coral}\small$\bullet$}]
    \textbf{\bodycolor{Coral}Mark $\mathcal{M}$.} 
    Marks constitute the fundamental graphical primitives that represent data items. 
    Formally, $\mathcal{M}=\{m_1, m_2, ..., m_n\}$ where $m_i$ is a mark type (\eg point, line, bar),
    which defines how data items are visually represented and the basic visual form of the chart.

    \item[{\bodycolor{teal}\small$\bullet$}]
    \textbf{\bodycolor{teal}Encoding $\mathcal{E}$.} This component specifies the mapping between data attributes and visual properties: $\mathcal{E} = \{(a_i, v_j)\}$ where $a_i$ is a data attribute and $v_j$ is a visual channel (\eg position, size, color). These mappings transform abstract data into perceptible visual forms.

    \item[{\bodycolor{MediumPurple}\small$\bullet$}]
    \textbf{\bodycolor{MediumPurple}Design $\mathcal{G}$.} The design component captures visualization properties not directly tied to data semantics, including $\mathcal{G} = \{g_1, g_2, ..., g_n\}$ where each $g_i$ represents stylistic elements like background color, gridlines, axis properties, chart title, etc. 
    These elements typically focus on the aesthetics and accessibility of the visualization.
\end{itemize}

\noindent
\textbf{Procedure.}
We analyze the performance of LLMs in interpreting natural language utterances to visualization specifications in two steps:
(1) \emph{LLM Processing} (\cref{fig:corpus-pipeline}), where we instruct the model with 814 utterances to generate visualizations through Vega-Lite specifications, followed by a self-explanation of its inference rationale and an accuracy evaluation;
(2) \emph{Expert Analysis}, where four domain experts \rev{(co-authors with extensive visualization and LLM experience)} independently analyze the results of LLM processing to identify the limitations of text prompts and the reasons behind LLM misinterpretation. 
Afterwards, the experts conduct a meeting to discuss and resolve any discrepancies in analysis results, addressing the two research questions.

\vspace{2pt}
\noindent
\textbf{Models.}
\rev{We benchmark four advanced LLMs (GPT-4o, Gemini-2.0-Flash, Gemini-3.0-Flash, and Claude-Sonnet-3.5) and observe consistent qualitative insights across models. We choose Gemini-2.0-Flash as the primary model to be reported because its performance is near the median among the four, making it a representative baseline for reporting detailed results, while a public visual analysis interface for each model's results is available at \url{https://wenzhen.site/vispilot/corpus}.}

\subsection{LLM Interpretation of Natural Language Utterances}
\label{sec:vis-spec}
To address \emph{\textbf{RQ1}}, we categorize the LLM interpretation patterns of natural language utterances into two types, depending on the explicitness of the specification expressed in the utterance:

\begin{itemize}[leftmargin=*]
    \item \textbf{{Explicit Reference}}: 
    The LLM interprets the user intent based on sufficient rationale and explicit specification of the visualization components.
    For example, \emph{"show me the average sales by region using a bar chart"} explicitly specifies the data schema {\bodycolor{CornflowerBlue}$\mathcal{D}=\{(sales, average), (region, raw)\}$} and the mark type {\bodycolor{Coral}$\mathcal{M}=\{bar\}$}.

    \item \textbf{{Implicit Inference}}:
    The LLM interprets the user intent based on implicit assumptions and incomplete specification of the visualization components.
    For example, \emph{"Show me the sales by region"} does not explicitly specify the aggregation function (\eg average, sum) or the mark type (\eg bar, line), leading to implicit inference of the data schema {\bodycolor{CornflowerBlue}$\mathcal{D}=\{(sales, sum), (region, raw)\}$}, mark type {\bodycolor{Coral}$\mathcal{M}=\{bar\}$}, and encoding {\bodycolor{teal}$\mathcal{E}=\{(region, x), (sales, y)\}$}.
\end{itemize}
We employ the self-explanation mechanism of LLMs~\cite{huang2023can} to identify these interpretation patterns.
The LLM is prompted to generate a self-explanation of its reasoning process, including the rationale for its interpretation with a classification of interpretation patterns (explicit or implicit) for all components in the generated specification.

\vspace{4pt}
\noindent
\textbf{Results.}
\hyperref[fig:nl-queries]{\Cref*{fig:nl-queries}-A} shows that the user utterances are widely ambiguous or incomplete, 
leading to at least three implicit inferences for each utterance ($\mu=7.27$, $\sigma=2.23$).
The top 3 frequent inference made by the LLMs is on the \designprop{mark style} (814 times), \encprop{x-axis encoding} (724), and \designprop{y-axis title} (567).
When the user requests the most complicated visualizations in the corpus, such as \emph{``For each country show the relationship between average acceleration and number of cylinders''}, the LLMs need to make 16 implicit inferences to generate the target visualization, involving \designprop{design} (10), \encprop{encoding} (4), \markprop{mark} (1), and \dataprop{data schema} (1) specifications. 
We then analyze the implicit inference patterns specifically on four specification components, as shown in \fref{fig:nl-queries}{B}.
\begin{itemize}[leftmargin=*]

\item 
\textbf{Data Schema (20\%).}
The data schema is the most frequently explicitly specified component, with only 20\% of utterances exhibiting ambiguity or incompleteness.
The most frequently inferred data schema is the \dataprop{y-axis aggregation} (151), which is often assumed to be \textit{sum} or \textit{average} when not explicitly stated.
Besides, the users also frequently omit the third \dataprop{data attribute} (139) in their utterances, which is often used for \textit{color} or \textit{column} in the encoding specification.

\item 
\textbf{Mark (43\%).}
More than half of utterances manage to explicitly specify the mark type through describing the chart type (\eg bar chart, line chart) or the shape of the marks (\eg point, line), while the other 36\% of utterances lead to implicit inferences on the mark type.
These utterances are often expressed in vague terms like \textit{``relationship''} or \textit{``compare''} for their analytic tasks, which do not explicitly indicate the mark type.
The most frequently inferred mark type is \markprop{bar} (47\% of 420 bar charts),
followed by \markprop{point/circle} (44\% of 262 scatter plots) and \markprop{line} (38\% of 95 line charts).

\item 
\textbf{Encoding (77\%).}
Most utterances lead to implicit inferences on encoding choices.
The users usually do not explicitly specify the encoding choices for coordinates (727 of \encprop{x-axis} and 509 of \encprop{y-axis}), which are often inferred by \textit{``by convention''} or \textit{``by experience''} as the LLMs explained.
The \encprop{color} encoding is merely inferred, as users often explicitly specify the color encoding such as \textit{``color by region''}.
            
\item 
\textbf{Design (96\%).}
The design component is the most frequently inferred specification, due to users rarely providing explicit design preferences in their utterances.
The LLMs typically infer the design properties based on their own preferences.
Notably, the Gemini model prefers to use default settings for \designprop{mark style} (814), \designprop{axis title} (567), and \designprop{axis format} (384), which often misalign with the target visualization.
When the utterance involves specific analytic intent, the model may also infer the design properties based on the context, such as \textit{``show me the distribution of average sales by region''} is inferred as a histogram with a binning design and a proper axis format.

\end{itemize}

\subsection{Analysis of LLM Misinterpretation}
\label{sec:text-limitation}
To address \emph{\textbf{RQ2}}, we evaluate the performance of LLMs on visualization generation by comparing the generated specifications against the ground truth.
We adopt hard constraints for the \dataprop{data schema}, \markprop{mark}, and \encprop{encoding} components, and soft constraints for the \designprop{design} component when evaluating the accuracy.
\rev{Specifically, the overall accuracy is defined as 1 if and only if all hard constraints strictly match the corresponding key-value pairs in the ground truth Vega-Lite specifications, otherwise 0,
while the bias on design components is not considered in the overall accuracy due to its subjectivity.}

\rev{\hyperref[fig:nl-queries]{\Cref*{fig:nl-queries}-C} shows an overview of four evaluated LLMs' accuracy for generating visualization specifications.}
The Gemini-2.0-Flash model achieves an overall accuracy of 47\% on the 814 utterances, with \dataprop{73\%} for {data schema}, \markprop{92\%} for {mark}, \encprop{69\%} for {encoding}.
We then investigate how these implicit inferences may lead to LLM misinterpretation and incorrect results. 
The accuracy is significantly lower when the specifications are implicitly inferred, with only \dataprop{53\%} for data schema, \markprop{85\%} for mark, and \encprop{71\%} for encoding.
\rev{Beyond accuracy metrics, four experts manually reviewed failure cases alongside the user utterance inputs to identify recurring intent-expression issues and misinterpretation patterns.
Through in-depth analysis and meetings, we identify three notable patterns as follows:}

\textbf{Limited Expression of Visual Intent.}
Text prompts are inherently inflexible and inconvenient to express users' visual intent. First, users usually use vague terms like \textit{``relation''} or \textit{``associate''}, rather than explicitly stating chart types like \textit{``line''} or \textit{``scatter''}. This typically yields much lower fidelity for marks, as LLMs might fail to capture users' nuanced intents. For example, \textit{``relationship between release year and average production budget''} leads to a bar chart instead of the expected line chart. Second, when dealing with encodings, text prompts often lack axis-specific details that clearly define the visual mappings and layout. A common misinterpretation by LLMs is the incorrect assignment of data attributes to the x-axis or y-axis, unless explicitly specified or certain phrases like \textit{``horizontal''} or \textit{``vertical''} are used. Additionally, when users' utterances involve a third encoding channel without providing explicit instructions, LLMs frequently struggle to infer the correct encoding type like color or size, especially in scatterplots or bubble charts. In summary, text prompts often fail to capture the full range of users' visual intent due to linguistic ambiguities and the difficulty of describing precise visual relationships through text alone.

\textbf{Inadequate Guidance of LLM Behavior.}
It is observed that, even when visualization specifications are explicitly stated in the utterances, LLMs still struggle to generate accurate visualizations.
In certain cases, the absence of specific keywords constantly leads to LLMs' incorrect inferences.
For example, keywords like \textit{``across''},  \textit{``over''}, and \textit{``by''} are very likely to guide LLMs to the correct encoding choices, while other terms like \textit{``between''} or \textit{``versus''} are less effective. However, users are often unaware of which keywords are essential for guiding LLMs' behavior, leading to misinterpretation. The situation is exacerbated when dealing with complex utterances that involve multiple encoding channels or uncommon designs. Even when users provide specific and detailed instructions, LLMs still struggle to generate accurate specifications. For example, we find that LLMs keep failing to create multi-view visualizations, such as small multiples or concatenated views, unless users explicitly mention relevant terms (\eg \textit{``separate''}, \textit{``split''}). A typical case is the misinterpretation of \textit{``the average of Production Budget categorized by Creative Type and parameterized by Content Rating''}, where LLMs always generate a single stacked bar chart instead of a faceted view. These observations indicate that users' lack of knowledge in formulating effective text prompts frequently hinders LLMs from interpreting and fulfilling their requirements.

\textbf{Misaligned Human-LLM Design Preferences.}
Visualization authoring relies on specific design knowledge drawn from established best practices. However, users' design preferences are often misaligned with LLMs' design choices learned from their training data. In our analysis, we find that LLMs tend to overwhelmingly favor certain default settings or common design patterns, leading to unexpected or inappropriate results. This misalignment is particularly evident in the misinterpretation of stylistic aspects like axis formats, color schemes, or chart titles. As users rarely provide explicit instructions for these visualization properties, LLMs resort to their preconceived notions, which may deviate from users' expectations, or sometimes decrease the readability or aesthetics of the visualization. For example, LLMs often generate scatterplots with hollow circles instead of filled ones, or use a sequential color scheme instead of a diverging one, which may not be suitable for the given data context or analytic tasks. Another common issue is the handling of time units, where LLMs frequently fail to infer the correct time granularity for the x-axis, causing visual clutter. These discrepancies between users' design preferences and LLMs' learned behaviors demand extensive customization operations and iterative enhancements, which can be frustrating and time-consuming.

\section{Leveraging Visual Prompt}

Informed by findings from the empirical study, we design a multimodal prompting framework that incorporates visual prompts to instruct LLMs for visualization authoring tasks.      

\subsection{Design Considerations}
\rev{We aim to bridge the communication gap between user intent and model interpretation by addressing the limitations of textual expression and design misalignment identified in our study (\cref{sec:text-limitation}).}
Therefore, our prompting framework is designed following four key design considerations:

\begin{itemize}[leftmargin=5.5mm]
       
    \item[\textbf{C1}] \textbf{Expressiveness}: The visual prompting should enable users to express their visualization intent in a flexible and comprehensive manner, addressing the \textit{limited expression of visual intent} present in text prompting approaches.

    \item[\textbf{C2}]  \textbf{Structured Interpretation}: Visual prompts should be systematically captured and translated into precise specification constraints that LLMs can accurately interpret, overcoming the \textit{inadequate guidance} issues and ambiguities inherent in text prompts.
    
    \item[\textbf{C3}]  \textbf{Interaction Intuitiveness}: Visual interactions should leverage users' natural understanding of visualization creation, enabling iterative refinement while bridging the \textit{design preference bias between human and LLMs} that often requires extensive customization in the text prompting approaches.

    \item[\textbf{C4}] \revadd{\textbf{Guided Grounding}: Visual prompts should be represented with visualization-specific grounding signals instead of pure image content. The framework should capture clear interaction intent and operation semantics, enabling LLMs to map visual input to executable specification constraints.}
  
\end{itemize}

\subsection{Visual Input Actions}

To support creating expressive visual prompts with flexible interactions (\textbf{C1}) \revadd{and provide grounding cues for LLM interpretation (\textbf{C4})}, we propose four fundamental visual input actions that can be performed on a sketch canvas. 
These actions are inspired by {a comprehensive literature review of existing visualization authoring tools}~\cite{chen2025interchat,shen2025prompting,Wang2023NLVis}, which highlights common user interactions in visualization authoring tasks.
As such, users can freely create expressive visual prompts in a sketch canvas environment using these actions:

\begin{itemize}[leftmargin=5.5mm]
    
\item[\icon{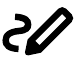}] \textbf{Scratch}: Users can sketch and layout visual elements on the canvas to represent their desired visualization. This action allows users to express their visual intent through freehand drawing, which can be interpreted as a specific mark type or layout.

\item[\icon{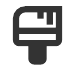}] \textbf{Style}: Users can apply visual styles to the marks, such as color, size, and shape, to convey their design or encoding preferences.
  
\item[\icon{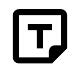}] \textbf{Annotation}: Users can annotate to the sketch visualization to provide additional context or information about specific elements.
\rev{The annotation act as deictic references that bind prompts directly to visual structures, eliminating the need for complex linguistic descriptions of spatial relationships.}
  
\item[\icon{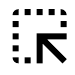}] \textbf{Manipulation}: Users can perform direct manipulation to select or modify visualization elements on the canvas, such as resizing or repositioning marks, to refine their design.

\end{itemize}

\begin{figure}[t]
    \centering
    \includegraphics[width=0.99\columnwidth]{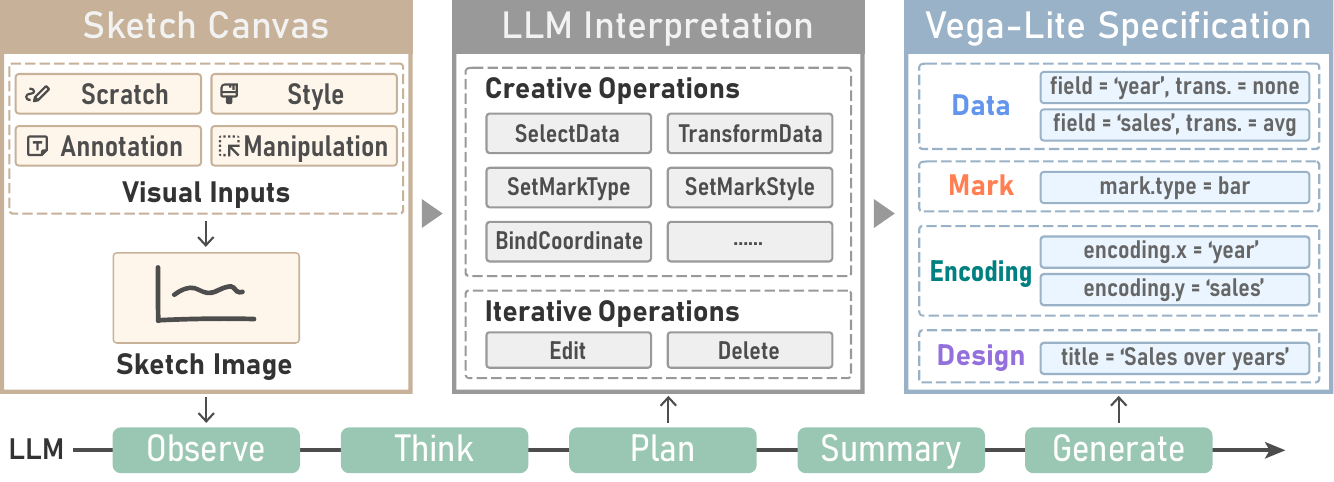}
    \caption{Our prompting framework instructs the LLM to interpret visual prompts to visualization specifications step by step.}
    \label{fig:visual-prompt}
    \vspace{-5mm}
\end{figure}

\subsection{Visualization Specification Operations}

To support LLMs in understanding visual input actions and accurately interpreting user intent (\textbf{C2}), we summarize the potential user intents as two categories of operations: {Creative Operations} and {Iterative Operations}.
The creative operations are derived from the specification components identified in our corpus analysis and are designed to be machine-interpretable.
Each operation represents the user intent on a specific aspect of the visualization specification.
Meanwhile, the iterative operations are designed to refine and enhance existing visualizations (\textbf{C3}).
To simplify the presentation, we include only the most common operations in this paper, while the complete list of operations is provided in the supplementary material.

\vspace{4pt}
\noindent\textbf{Creative Operations.} The creative operations aim to create a new visualization from scratch, allowing users to express their visual intent covering the entire specification space.

\begin{itemize}[leftmargin=*]
    \item \textbf{SelectData} specifies the data attributes to be visualized. 
    Users can indicate their selection of attributes by \icon{note} annotate keywords in the chart title, legend, or axis labels.
    This operation represents constraints on the \dataprop{field} properties in the data schema specification.

    \item \textbf{TransformData} specifies the data transformations to be applied to the selected data attributes.
    Users can express the transformation requirements through explicit \icon{note} annotations along with data attributes or \icon{scratch} visual examples, for instance, sketching a descending bar chart to indicate sorting.
    This constrains a series of properties in the data schema specification, including \dataprop{aggregate}, \dataprop{sort}, and \dataprop{transform}, etc.

    \item \textbf{SetMarkType} specifies the type of mark to be used in the visualization.
    Users can \icon{scratch} sketch desired visual marks on the canvas, which presents constraint on the \markprop{type} property in the mark specification.

    \item \textbf{SetMarkStyle} specifies the visual styles to be applied to the marks.
    Users can \icon{style} apply styles to the marks, such as color, size, and opacity, to convey their design or encoding preferences.
    This constrains on either properties in the design specification such as \designprop{mark.fill},
    or properties in the encoding specification such as \encprop{encoding.color}.

    \item \textbf{BindCoordinate} specifies the mapping of data attributes to coordinate axes.
    Users can \icon{scratch} sketch the coordinate axes with data attributes \icon{note} labeled on them to indicate their intents.
    This typically constrains properties like \encprop{x.field}, \encprop{y.field} in the encoding specification.

    \item \textbf{Layout} specifies the layout of the visualization views.
    Users can \icon{scratch} sketch the layout of the visualization views on the canvas, such as arranging multiple charts in a grid or juxtaposition.
    This constrains properties such as \encprop{column.field} or \encprop{facet} in the encoding specification.
\end{itemize}

\vspace{4pt}
\noindent\textbf{Iterative Operations.} The iterative operations refine and enhance an existing visualization through progressive refinement.

\begin{itemize}[leftmargin=*]
    \item \textbf{Edit} modifies existing visualization specifications to refine the visualization.
    Users can \icon{selection} select elements to highlight their target with visual expressions, such as changing the color of marks or modifying the axis title.
    This instructs the model to edit the corresponding visual elements on the basis of the existing visualization specification.
    
    \item \textbf{Delete} removes specific elements or properties from the existing visualization.
    Users can \icon{selection} select elements to remove or properties to disable.
    This sets specific properties to null/false or removes them entirely from the Vega-Lite specification.
\end{itemize}

\subsection{Interpretation of Visual Prompts}
\label{sec:prompting}
To instruct LLMs to accurately interpret visual prompts, we propose a novel multimodal prompting framework that systematically guides the model through visual input interpretation and visualization generation in a step-wise reasoning manner.

\vspace{4pt}
\noindent
\textbf{Framework.}
Our framework structures the LLM reasoning process through five sequential steps, as illustrated in Figure~\ref{fig:visual-prompt}.
The model reasoning starts with \emph{Observation}, where the model receives visual inputs and observes user actions on the sketch canvas.
The model is required to report its observations with specific location information (\eg \emph{``the user annotated `year' on the x-axis''}) to ensure it accurately captures user actions in a visual context.
Next, the model proceeds to \emph{Thinking}, where it infers the user's underlying visualization intent based on the observed visual input context and the conversation history, identifying which specification components (data, mark, encoding, etc.) the user is trying to express.
After that, the model enters the \emph{Planning} phase, where it maps the inferred intent to concrete specification operations (SelectData, SetMarkType, BindCoordinate, etc.) and generates a list of operations to be performed.
Subsequently, the model generates a \emph{Summary} of its understanding and planned operations, explicitly stating its interpretation of the user's intent and the operations it will perform to fulfill that intent.
Finally, the reasoning process ends with the \emph{Generation} phase, where the model produces a complete Vega-Lite specification that implements all the visualization requirements expressed through the visual prompts.

\vspace{4pt}
\noindent
\rev{\textbf{Technical Implementation.}
We operationalize this structured reasoning process through a comprehensive system prompt (please refer to supplementary materials for details).
To support the iterative human-in-the-loop workflow, the prompting framework aggregates a composite input context for each turn: (1)\emph{conversation history} for semantic continuity; (2) \emph{data context}, comprising the dataset profile, headers, and sample rows for precise data binding; and (3) \emph{visual context}, combining the rasterized canvas image with textual descriptions of user actions (e.g., element selections) that are not discernible from the snapshot. 
Finally, the model is instructed to output reasoning rationales and the Vega-Lite specification in a unified JSON format to facilitate automated parsing.}

\begin{figure*}[t]
    \centering
    \includegraphics[width=0.99\textwidth]{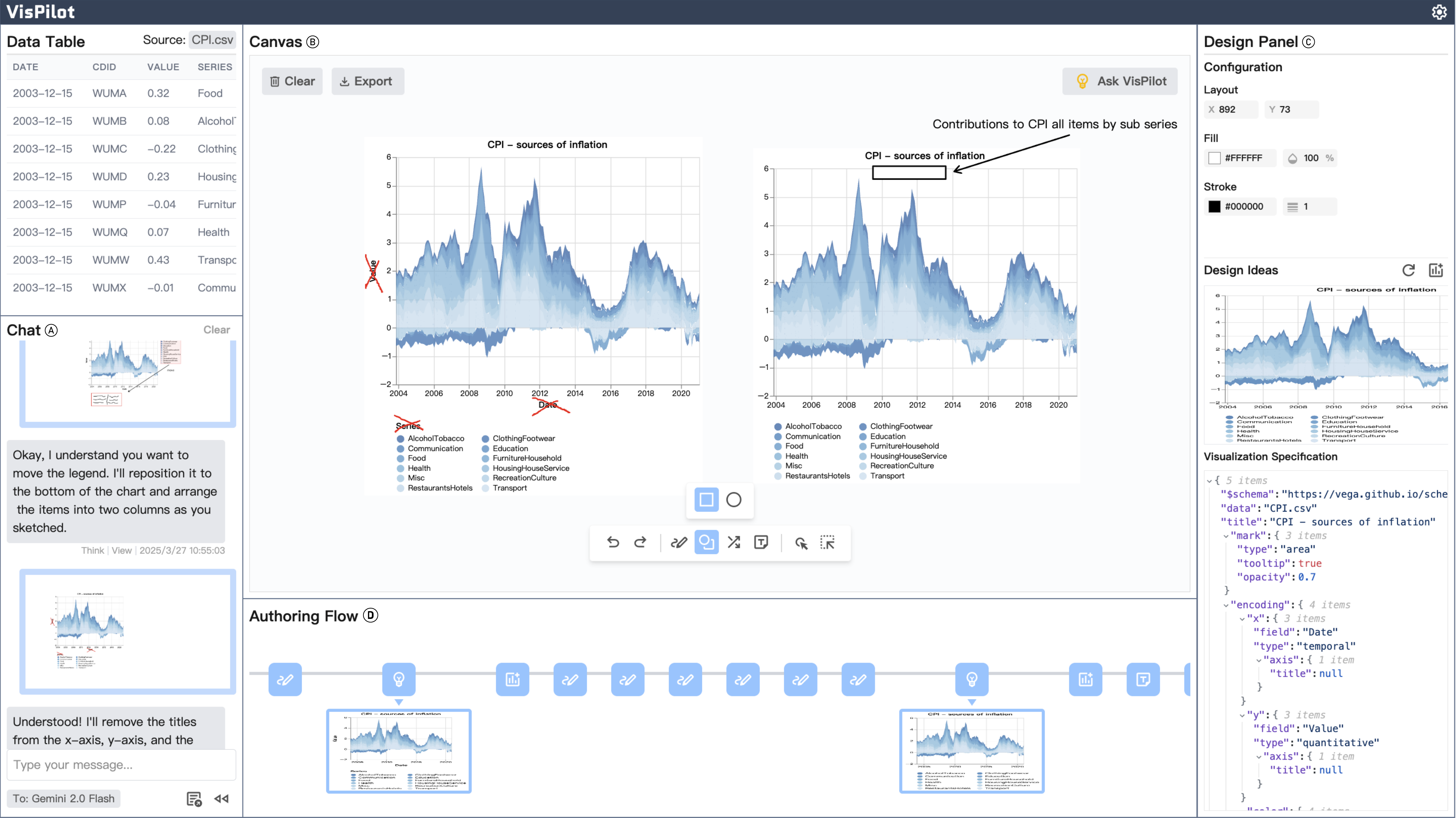}
    \caption{The interface of VisPilot includes four components: (A) Chat Interface, (B) Free-drawing Canvas, (C) Design Panel, and (D) Authoring Flow.}
    \label{fig:vispilot-interface}
    \vspace{-5mm}
\end{figure*}

\section{The VisPilot System}
To instantiate the guided visual prompting framework described above, we design VisPilot as a proof-of-concept system.
\revadd{VisPilot preserves the flexibility of sketch-based interaction, as explored by systems such as Code Shaping~\cite{Yen2025CodeShaping}, while employing structured user interface to make visualization-specific intents easier to express. 
The interface components act as lightweight guidance that help users compose spatial references, visual designs, and iterative edits without relying solely on free-form sketching or verbose text descriptions.}

\subsection{Interface Design}

The interface of VisPilot consists of four main components: a chat interface, a free-drawing canvas, a design panel, and an authoring flow view (as shown in \Cref{fig:vispilot-interface}).

\textbf{Chat Interface.}
The chat interface comprises a data table view and a chat view (\fref{fig:vispilot-interface}{A}).
Users can upload a dataset to start the conversation with the system.
The dataset is displayed in the data table view and automatically sent to the LLM.
Below the data table, the LLM initially responds to the dataset and ask the user for a specific visualization task.
In the following conversation, users can instruct the LLM to generate visualizations by natural language in this view.
As we employ the prompting framework presented in \cref{sec:prompting}, the LLM commonly generates long responses containing its thinking process and visualization specification code.
We hide these long responses and only show the summary part of responses to avoid overwhelming users with too much information.

\textbf{Free-drawing Canvas.}
The free-drawing canvas allows users to create visual prompts for the LLM (\fref{fig:vispilot-interface}{B}).
Users can use mouse, touch, or pen as input devices to draw on the canvas.
Refer to the prompting framework (\cref{sec:prompting}), we provide a set of widgets to help users create visual prompts for the LLM.
The pen, shape, axis, and text widgets are used to create sketches of visualizations, while the two selection widgets are used to select the elements drawn by users or generated by the LLM for further manipulation.
Once the user clicks on the Ask VisPilot button, the system will send the sketch image as a visual prompt to the LLM.

\textbf{Design Panel.}
The design panel presents a configuration panel and a design idea panel (\fref{fig:vispilot-interface}{C}).
The configuration panel allows users to configure the style properties of selected elements in the canvas.
The design ideas panel displays the design ideas generated by the LLM based on user instructions, including the Vega-Lite specification code and the corresponding visualizations.
Users can add the satisfied visualizations to the canvas or ask the LLM to generate alternative design ideas.

\textbf{Authoring Flow.}
The authoring flow of VisPilot is shown in \fref{fig:vispilot-interface}{D}.
It shows the steps of the authoring process made by the user.
All user interactions and LLM responses are recorded in the authoring flow.
The interactions for text prompting and visual prompting are shown in gray and blue colors, respectively.
Each visualization generated by the LLM is shown below the timeline, presenting the process of how the user completed the authoring task.

VisPilot is implemented as a web application using React and Vega-Lite at frontend, with generative capability supported by OpenAI-compatible API.
The system is accessible through desktop and tablet devices, allowing users to input multimodal prompts using the mouse, touch, pen, or keyboard.

\subsection{Usage Scenario}
\rev{To demonstrate the effectiveness of VisPilot, we present a usage scenario that showcase how users can leverage multimodal prompts to complete visualization authoring tasks.}

\label{sec:case-1}
The usage scenario follows Alice, a data journalist, in her authoring process of a visualization showing how different categories of consumer goods contribute to inflation over time (\cref{fig:teaser}). 
The process begins with Alice uploading a dataset containing the Consumer Price Index (CPI) data from 2003 to 2021.
In the data table view, she can see the dataset including the CPI values for various categories of goods and services, such as food, education, and transportation.
Afterwards, she has an idea of creating a streamgraph-style area chart to visualize the contribution of different categories to the overall inflation rate, which she sketches on the free-drawing canvas (\hfref{fig:teaser}{A1}).
VisPilot then generates a visualization based on her sketch, displaying the contribution of different categories to the overall inflation rate over time (\hfref{fig:teaser}{B1}).

Alice is satisfied with the visual form of the generated chart but does not like the default color scheme, which she finds visually complex.
To address this, she modifies her sketch by drawing a colored example on the right side of the chart, indicating her preference for a gradient blue color palette (\hfref{fig:teaser}{A2}).
VisPilot responds by implementing a sequential blue color scheme in the chart (\hfref{fig:teaser}{B2}).
To further enhance the readability of the chart, Alice refines the color scheme by reversing example colors, and annotates titles on the sketch (\hfref{fig:teaser}{A3}).
After the system generates the updated chart with the new color scheme and titles (\hfref{fig:teaser}{B3}), Alice is pleased with the overall design but wants to make some final adjustments directly on the visualization.

She uses the selection tool to select the legend and draws an arrow pointing to her desired position (\hfref{fig:teaser}{A4}), prompting the system to reposition the legend to the top-right area of the plot (\hfref{fig:teaser}{B4}).
She further simplifies the design by crossing out the axis titles (\hfref{fig:teaser}{A5}), which the system removes from the chart (\hfref{fig:teaser}{B5}).
As a complementary step, Alice decides to add a statement under the chart title to provide context for the visualization (\hfref{fig:teaser}{A6}).
VisPilot interprets her prompts and adds the statement as the subtitle of the chart (\hfref{fig:teaser}{B6}).
Finally, Alice is satisfied with the final design and saves the visualization.
This case study demonstrates that our prompting framework is effective in guiding the LLM to generate visualizations that precisely align with user preferences.

\begin{figure*}[t]
    \centering
    \includegraphics[width=0.99\textwidth]{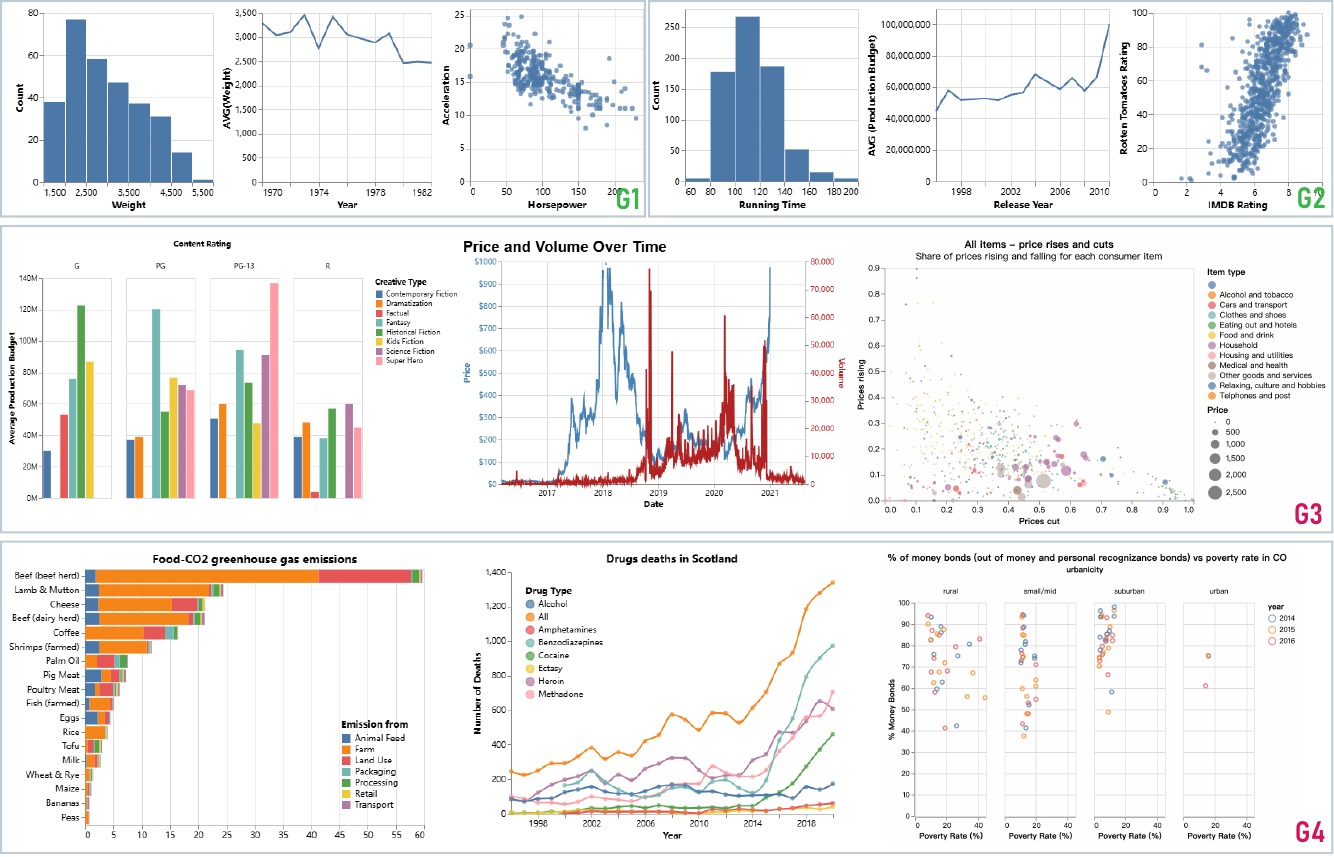}
    \caption{The target visualizations of tasks. Participants need to complete one group of simple tasks (\textbf{\bodycolor{MediumSeaGreen}G1}, \textbf{\bodycolor{MediumSeaGreen}G2}) and one group of complex tasks (\textbf{\bodycolor{MediumVioletRed}G3}, \textbf{\bodycolor{MediumVioletRed}G4}).}
    \label{fig:evaluation-target}
    \vspace{-5mm}
\end{figure*}

\section{User Study}
\label{sec:user-study}
We conducted a controlled user study to evaluate the effectiveness of multimodal prompting for visualization authoring.

\subsection{Methodology}
\rev{VisPilot targets users with varying levels of visualization knowledge, ranging from novices to professionals.}
Our study followed a within-subject design to account for the individual differences among participants.
VisPilot was used as the technology probe to explore the potential of multimodal prompting for visualization authoring.
We aim to evaluate the multimodal prompting approach from three aspects: (1) the accuracy and efficiency of creating visualizations, (2) the usability and user experience of the system, and (3) the user behavior and interaction patterns in different prompting conditions.

\vspace{4pt}
\noindent
\textbf{Conditions.}
The study was conducted in two conditions:

\begin{itemize}[leftmargin=*]
    \item \textbf{Text Condition (baseline).}
    In the text prompting condition, users can interact with the system only through the chat interface, which provides similar user experience to common LLM chat interfaces (\eg ChatGPT).
    Users can type their natural language utterances, and the system responds with chat messages and visualizations.
    \rev{This condition represents the standard interaction paradigm closely aligned with realistic usage scenarios of LLMs for visualization authoring.}
    
    \item \textbf{Multimodal Condition.}
    In the multimodal prompting condition, users can interact with the system without restrictions.
    Comparing to the text condition, users have additional capabilities to create visual prompts from scratch or existing visualizations.
    They are free to use text or multimodal prompts to convey their intents, and the system respond with chat messages and visualizations as the text condition.
\end{itemize}
The two conditions both used the gemini-2.0-flash-001 model as the LLM backend, along with the same hyper-parameter configuration and system prompt.

\vspace{4pt}
\noindent
\textbf{Participants.}
\revadd{We recruited 15 participants (aged 21--31; 5 female and 10 male).}
\rev{All participants reported prior experience using LLMs for data analysis and visualization authoring, with at least one year of regular use.}
\rev{Two participants (P1--P2) were professionals in a fintech company whose daily work involves data analysis and visualization.}
\rev{The remaining participants (P3--P15) were graduate or undergraduate students from a local university, who commonly need to create visualizations for data analysis and research.}
\rev{Participants had varying levels of visualization authoring experience, ranging from visual tools (Excel, Tableau), visualization grammars (Vega-Lite) to programming libraries (Matplotlib, D3.js, etc.).}

\vspace{4pt}
\noindent
\textbf{Tasks.}
We designed a set of replication tasks in the study.
\rev{The target visualizations were selected from two Vega-Lite visualization datasets~\cite{Arjun2021Collecting,Ko2024Dataset}.}
There were four groups of visualizations for the participants to replicate (see \cref{fig:evaluation-target}), including two simple-level groups (\textbf{\bodycolor{MediumSeaGreen}G1}, \textbf{\bodycolor{MediumSeaGreen}G2}), and two complex-level groups (\textbf{\bodycolor{MediumVioletRed}G3}, \textbf{\bodycolor{MediumVioletRed}G4}).
\rev{The complexity level was categorized by the number of specification items required to determine the visualization.
The simple-level tasks required 7--8 specification items and involved exactly 2 encoding channels, while the complex-level tasks required 10--15 specification items and involved more than 2 encoding channels.}
Each group was designed to contain three types of visualizations including bar chart, line chart, and scatter plot.
Participants were asked to replicate one simple and one complex group of visualizations in each condition (12 trials in total).
They were asked to create visualizations as similar as possible to the targets,
including the data schema, marks, visual encodings, and design details.

\vspace{4pt}
\noindent
\textbf{Procedure.}
We conducted a 90-minute session with each participant. After obtaining informed consent, participants completed a pre-study questionnaire about their experience with data visualization tools and LLM-based systems. 
We then provided a 10-minute introduction to VisPilot, including a demonstration of its features in two prompting conditions. 

The main study consisted of four sessions where participants replicated visualization groups using VisPilot. 
We counterbalanced the order of conditions across participants to mitigate learning effects. 
Each participant experienced both conditions (text-only and multimodal) and both complexity levels (simple and complex). 
For each session, participants were given the target visualization and instructed to recreate it as accurately as possible within a 10-minute time limit. 
We encouraged participants to think aloud during the process.

After completing all trials, participants filled out a post-study questionnaire to evaluate the system usability and cognitive load. We concluded with a semi-structured interview to gather qualitative feedback about their experiences using different prompting modalities. All sessions were recorded for later analysis with screen recordings.
Each participant was equally compensated with \$20 for their time and effort.

\vspace{4pt}
\noindent
\textbf{Metrics.}
We measured two conditions in terms of quantitative and qualitative metrics.
The quantitative metrics evaluated the task performance including efficiency and accuracy.
For task efficiency, we recorded the \emph{completion time} at two time points, including the first-time creation of the visualization and the last-time generation by the LLM, which represent the time for completing creation and iteration tasks, respectively.
For task accuracy, we recorded the \emph{number of mismatches} at two time points as above, which represents the number of mismatched visualization properties. 
Qualitative metrics evaluated system usability and user experience through 7-point Likert scale questionnaires, collected from post-study interviews.

\subsection{Results}
We analyzed the data from the study, including quantitative metrics, qualitative feedback, and user behavior patterns.

\subsubsection{Quantitative Results}
Overall, the multimodal condition achieved higher task accuracy without significantly affecting task efficiency compared to the text condition.
\revadd{We first applied Shapiro-Wilk tests to the participant-level paired differences between two conditions. For comparisons whose paired differences did not significantly deviate from normality (\(p \ge .05\)), we used paired t-tests; otherwise, we used two-sided Wilcoxon signed-rank tests.}
The details of the results are as follows:

\vspace{2pt}
\noindent
\textbf{Task Efficiency.}
\revadd{The multimodal condition did not significantly differ from the text condition in completion time for first-time creation (Simple: $p=.181$; Complex: $p=.145$) or last-time generation (Simple: $p=1.000$; Complex: $p=.570$).} This suggests that, even typing text prompts was generally faster than drawing visual prompts, the effort of visual interaction did not significantly affect the overall task efficiency. During the study sessions, we constantly observed that, in the text condition, participants tended to spend more time on thinking and formulating the prompts than on the actual interaction with the system, while vice versa in the multimodal condition. This can be ascribed to the intuitiveness and directness of visual prompts in the multimodal condition, which allowed participants to quickly iterate on their visualizations, thereby complementing the time spent on drawing and manipulating visual elements.

\vspace{2pt}
\noindent
\textbf{Task Accuracy.}
\revadd{The multimodal condition achieved fewer mismatches in total compared to the text condition for the first-time creation of visualizations (Simple: \(\Sigma=55<62\); Complex: \(\Sigma=181<232\)) and the last-time generation by the LLM (Simple: \(\Sigma=0<2\); Complex: \(\Sigma=6<14\)). For first-time creation, the multimodal condition was significantly better for total mismatches in complex tasks ($p=.014$), especially for data schema, marks, and encoding properties ($p=.026$), while the difference was not significant for total mismatches in simple tasks ($p=.103$).}
Generally, multimodal prompts helped participants obtain visualizations that were more aligned with their expectations. 
We noticed that for subtle design properties such as axis ticks/formats, color schemes, and legend placements/orientations, using visual prompts was more precise and user-friendly than pure texts.
This demonstrates the superiority of multimodal prompting in conveying detailed and nuanced visual intents and reducing users' cognitive load, especially for complex visualizations that exhibit uncommon encoding choices or require specific spatial and layout considerations.

\begin{figure}[b]
    \centering
    \includegraphics[width=0.99\columnwidth]{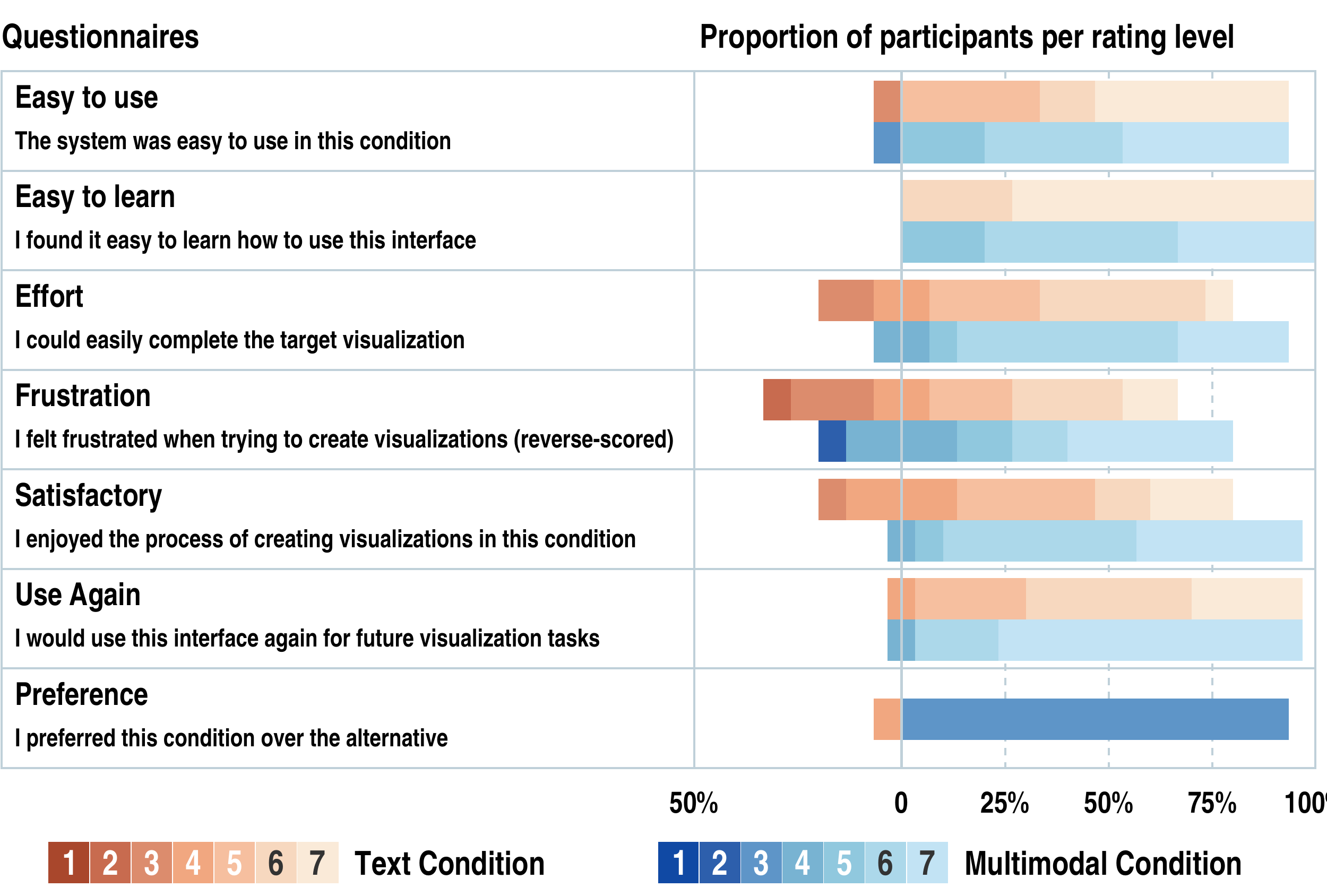}
    \caption{The questionnaire and qualitative results from \revadd{15 participants}, where rating from 1 to 7 represents \textit{strongly disagree} to \textit{strongly agree}.}
    \label{fig:eval-res}
    \vspace{-5mm}
\end{figure}

\subsubsection{System Usability}
The usability results are summarized in \cref{fig:eval-res}. Overall, participants appreciated the usability of the multimodal condition for creating visualizations. We analyze the results from the following dimensions:

\vspace{2pt}
\noindent
\textbf{Easy to Use and Learn.}
\revadd{The multimodal condition achieved a comparable score to the text condition in terms of ease of use (\(\mu=6.0, \sigma=1.1\)) and learnability (\(\mu=6.1, \sigma=0.7\)).} The reason why the multimodal condition did not score higher was that, text prompts were inherently intuitive and easy to use for participants who were already familiar with LLMs. Nevertheless, it did not require a steep learning curve to use our proposed visual input actions, as stated by most participants. Once they became accustomed to the interface, they found it \textit{``natural and intuitive''} (P3), just like \textit{``drawing on a whiteboard''} (P5). Also, some participants mentioned that the multimodal condition would be \textit{``more accessible to people not familiar with LLMs''} (P2) and \textit{``even more useful with a tablet for students to learn about visualizations''} (P2).

\vspace{2pt}
\noindent
\textbf{Effort and Frustration.}
\revadd{The multimodal condition scored higher in terms of effort (\(\mu=5.9, \sigma=1.0\)) and frustration (\(\mu=5.5, \sigma=1.6\)) compared to the text condition.} We observed that participants struggled with the text condition when creating visualizations that required spatial and layout considerations. 
As stated by P8 \textit{``I could not move the legend to the bottom-right corner inside the chart, no matter how hard I tried to rephrase my text prompts''}. 
This significantly increased their effort and frustration levels. In contrast, the multimodal condition allowed participants to directly add an arrow on the canvas to indicate the desired position, which was more \textit{``effective and convenient''} (P9). However, P6, \rev{who was experienced with the Vega-Lite grammar}, found the multimodal condition \textit{``less efficient for simple tasks''}, as he could proficiently write accurate instructions rather than tediously drawing them. This suggests that different users may have different preferences for the two prompting modalities based on their domain expertise.

\vspace{2pt}
\noindent
\textbf{Satisfaction and Real-world Application.}
\revadd{The multimodal condition was more favored than the text condition in terms of satisfaction (\(\mu=6.2, \sigma=0.9\)) and willingness to use again (\(\mu=6.6, \sigma=0.8\)).} 
For example, P2, remarked, \textit{``visual interactions were more engaging and fun to use, making me feel like I was designing visualizations rather than just typing text prompts''}. 
They also suggested several features to improve the multimodal interface, such as \textit{``supporting pre-defined sketch templates''} (P9) and \textit{``enabling more precise selection and control of visual elements''} (P5). 
Additionally, most participants were willing to integrate visual prompts into their daily workflow for creating visualizations. 
They believed that visual inputs could be \textit{``a great complement to existing text-based visualization authoring tools''} (P10) and could potentially be applied to various scenarios, such as \textit{``visualization education''} (P2), \textit{``collaborative design sessions''} (P6), and \rev{\textit{``exploratory data analysis''}, as suggested by a fintech professional (P1)}.

\vspace{2pt}
\noindent
\textbf{Preference.}
\revadd{Most participants (14/15) expressed a preference for the multimodal condition over the text condition.}
While text prompts are concise and intuitive for simple user requirements, participants acknowledged their inherent limitations in conveying nuanced visual intents, especially for complex visualizations and detailed visual changes. 
Also, as stated by \rev{a participant majored in design science (P9)}, the multimodal condition was particularly useful for \textit{``creative scenarios where the desired outcome might be difficult to articulate with words alone''}. 
Meanwhile, some participants (P3, P8) wished to combine visual input with text input to \textit{``make the most of each other's advantages''} (P3). Regarding the only one participant (P7) who preferred the text condition, he explained that, \textit{``for me, text prompts are more efficient to express my intents, for example, typing the word `sort' is far more convenient compared to drawing multiple bars with lengths that decrease in size''}. We attribute this to different user habits, reinforcing the importance of providing multiple prompting modalities to cater for various user needs and preferences.

\vspace{2pt}
\subsubsection{User Behaviors and Interaction Patterns}

We observed notable user behavior patterns across the two prompting conditions, beyond the quantitative and qualitative metrics.

\vspace{2pt}
\noindent
\textbf{Text Prompting Strategies.}
In the text condition, participants often engaged in iterative prompt refinement: when initial outputs did not match their intent, they would rephrase, simplify, or break down their requests into smaller steps. For instance, P8 described a process of {``trial and error with wording''} to achieve the desired legend placement, but still found it difficult to control spatial details.

\vspace{2pt}
\noindent
\textbf{Multimodal Iteration Patterns.}
In the multimodal condition, participants frequently adopted a ``sketch-and-adjust'' workflow: they would first sketch the overall layout visually, then use direct manipulation to fine-tune elements, and occasionally supplement with text for precise specifications. This allowed for rapid, incremental adjustments and immediate feedback, which participants found intuitive and engaging.

\vspace{2pt}
\noindent
\textbf{Cognitive Load and Focus.}
We observed that in the text condition, participants spent more time planning and formulating prompts, while in the multimodal condition, their attention shifted to manipulating visual elements and interpreting system feedback. Several participants (e.g., P3, P5) noted that visual interaction \emph{``felt more like designing''} and reduced the need to mentally translate visual ideas into words.

\vspace{2pt}
\noindent
\textbf{Emergent Best Practices.}
Some participants developed hybrid strategies, such as using visual prompts for layout and text for data mapping, sketching with text annotation, suggesting the value of hybrid modalities. 
\rev{We observed that participants with more experience in Vega-Lite grammars used hybrid interaction more proactively and fluently. 
They tended to rely more on concise text instructions for simple or well-defined changes, while participants with less formal visualization knowledge benefited more from visual prompts for communicating spatial intents without needing specialized terminology.}

These observations reveal that multimodal prompting not only improves efficiency for complex tasks, but also supports more natural and flexible authoring behaviors.

\section{Expert Evaluation}
\rev{To evaluate how the system supports open-ended and creative visualization authoring in realistic professional workflows, we invited two domain experts to perform open-ended authoring tasks, followed by semi-structured interviews to gather in-depth qualitative feedback.}

\subsection{Case Study}

\rev{We recruited two experts with over five years of experience in data analysis and visualization.
E1 is a football analytics blogger and content creator, while E2 is a senior data journalist experienced in creating data-driven stories. 
Both two experts are proficient in traditional tools, e.g., Tableau, Python/Pandas and Excel, but had no prior exposure to VisPilot. 
The experiment lasted approximately 90 minutes.
We first introduced the study background and system features to the experts like user study.
The study utilized the English Premier League players statistics during the 2024/25 season as the dataset, which contains a rich set of 53 performance metrics for 562 players, such as Goals, Passes, Tackles, Position, etc.
The experts were given an open-ended objective: \emph{``Find outstanding attacking defenders and demonstrate their strengths through visualizations.''}
They were encouraged to think aloud and freely combine text and visual prompts to achieve their goals.}

\begin{figure*}[t]
    \centering
    \includegraphics[width=0.99\textwidth]{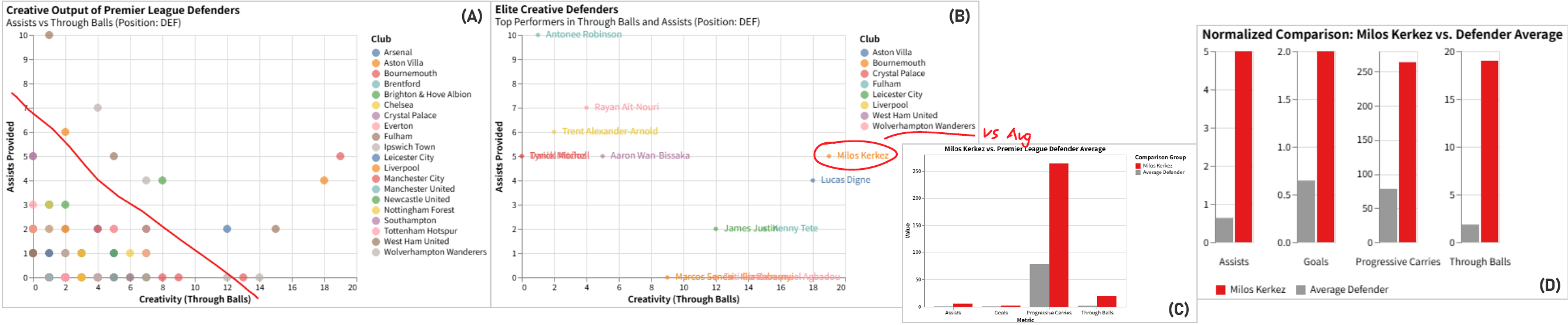}
    \caption{Case study with domain experts analyzing Premier League player statistics, using VisPilot to iteratively create visualizations from (A) to (D).}
    \label{fig:case-study}
\end{figure*}

\rev{\textbf{Broad Exploration with Vague Intent.} 
The experts started with a high-level intent to see which defenders contribute most to the attack. 
They initiated the task by directly typing \emph{``outstanding attacking defenders''} in the chat interface.
The system responded with a scatterplot of Goals vs. Assists. 
While factually correct, the experts noted that this view was too generic and suffered from overplotting due to the discrete integer values of goals and assists. 
To dig deeper into playmaking style rather than just scoring output, they decided to redefine the analysis dimensions. 
They scratched out the x-axis titles and scribbled \emph{``Creativity''} below it.
VisPilot interpreted this high-level concept by mapping it to \emph{``Through Balls''}, a fitting metric for defender creativity, and updated the chart to ``Assists vs. Through Balls'' (\fref{fig:case-study}{A}).}

\rev{\textbf{Spatial Filtering via Sketching.} 
The updated chart revealed a spread of players, but the experts wanted to strictly isolate the elite performers from the average. 
Instead of formulating a complex query (e.g., ``Filter for assists $\ge$ 4 OR through balls $\ge$ 8''), they simply drew a diagonal line across the canvas, visually separating the dense cluster of average players in the bottom-left from the outliers in the top-right. 
VisPilot interpreted this spatial divider as a filtering constraint. 
It zoomed in on the ``Elite'' region, removing the low-performing data points, and automatically labeled the remaining players to aid identification (\fref{fig:case-study}{B}).}

\rev{\textbf{Comparative Layout Design.} 
To further demonstrate the unique strengths of specific players, the experts decided to create a composite dashboard view. 
They focused on Milos Kerkez, whose position in the scatterplot suggested high creativity with medium assists. 
They circled the data point for Kerkez and drew a extending line pointing to the empty space on the right side of the canvas. 
At the end of the line, they sketched a blank box and annotated \emph{``vs Avg''}. 
VisPilot interpreted this visual linkage as a request for a comparison view targeting the selected entity. It generated a bar chart contrasting Kerkez against the league average defender (\fref{fig:case-study}{C}), notably augmenting the data by adding {Progressive Carries} to the chart—a key metric for Kerkez that was not present in the initial view.}

\rev{\textbf{Stylistic Refinement.} 
The experts were satisfied with the metrics but noticed a visual scaling issue: the high value of Progressive Carries ($\sim$260) dwarfed the other metrics ($<$20). To fix this, they circled the small-size bars and annotated \emph{``scale up''}. 
VisPilot recognized the readability issue and transformed the grouped bar chart into a faceted chart with independent y-axes (\fref{fig:case-study}{D}), effectively bridging the gap between exploration and publication-ready graphics.}

\subsection{Expert Interview}
\rev{Following the case study, we conducted semi-structured interviews to gather experts' feedback on how VisPilot integrates into professional workflows.
The experts responded positively to the system, particularly highlighting its ability to handle real-world ambiguity and support fluid analytical thinking. Their feedback is summarized as follows:}

\rev{\textbf{Ambiguity Resolution.}
Experts highlighted that visual prompts resolve ambiguity not only semantically, but also structurally. 
While the LLMs can infer abstract concepts from text like ``Creativity'', E2 noted that using text prompts sometimes result in unwanted layout shifts (e.g., regenerating a completely new chart type). 
In contrast, scratching out the axis and annotating in situ provided explicit visual grounding. 
As E2 explained, \emph{``Text is open-ended, which is risky for design stability. Writing on the canvas feels like applying a constraint: `Keep everything else, just change this part.' It gives me the control that conversational interfaces usually lack.''}}

\rev{\textbf{Cognitive Offloading.}
A recurring theme was the efficiency of visual prompts for spatial logic. 
E1 highlighted the friction of text-only interfaces for outlier detection: \emph{``Describing that specific diagonal boundary in text would require me to check axis ranges and write a complex SQL-like condition. Drawing a line is instantaneous. It keeps me in the `analysis flow' without breaking into coding logic.''}}

\rev{\textbf{Creative Brainstorming.}
Beyond efficiency, the multimodal interface lowers the barrier for expanding the analysis. 
E2 noted that standard visualization tools often require a rigid ``mode switch'' to create linked views, e.g., opening new sheets, configuring filter actions. 
In contrast, the sketching interaction allowed for fluid experimentation. 
\emph{``Usually, building a comparative view is a separate commitment that interrupts analysis,''} E2 remarked. 
\emph{``Here, drawing an extending line felt like a natural continuation of my thought process—I could inspect a hunch about a player instantly without having to navigate menus or configure a dashboard layout.''}}

\begin{figure*}[t]
  \centering
  \includegraphics[width=\textwidth]{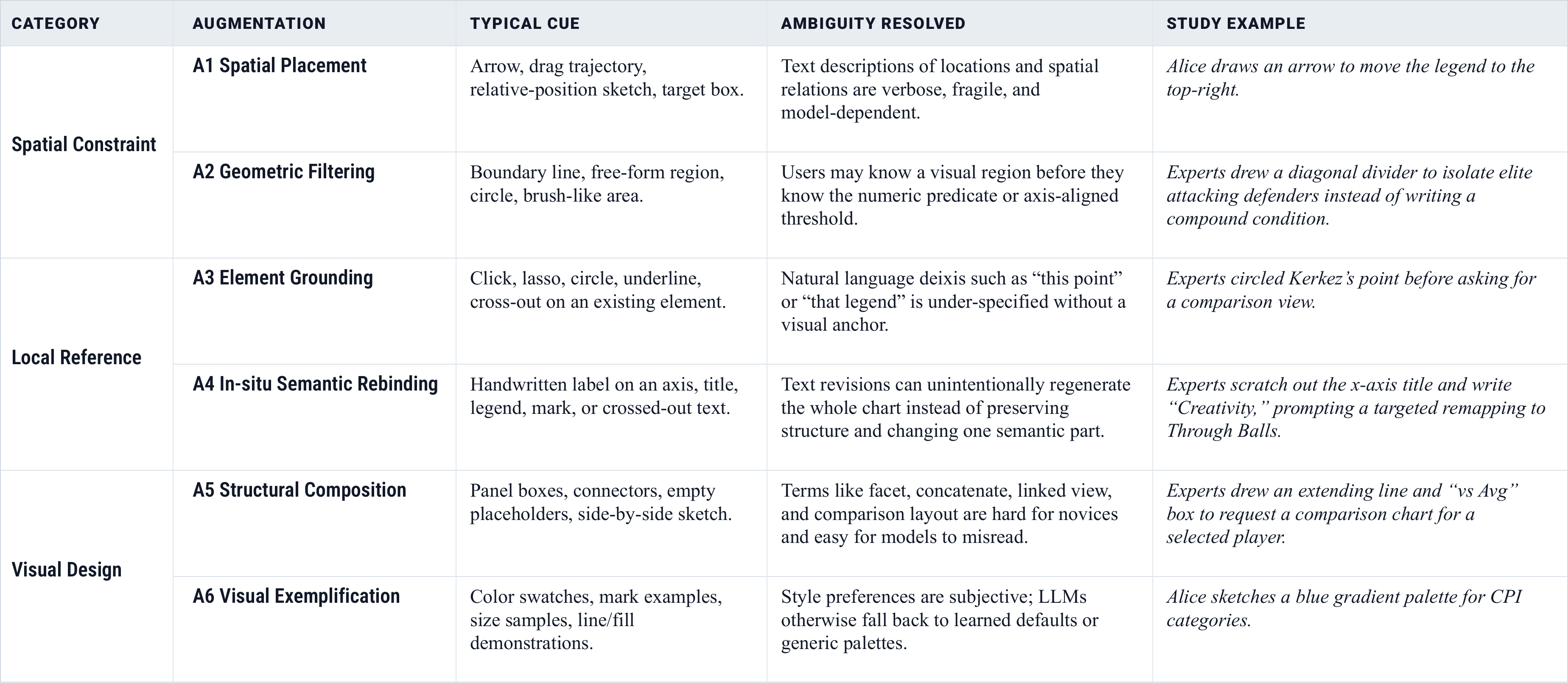}
    \caption{A taxonomy of multimodal prompting augmentations for visualization authoring. Each type summarizes a recurring way in which visual prompts complement text prompts by grounding user intent in visualization objects, spaces, structures, or examples.}
  \label{fig:prompting-taxonomy}
\end{figure*}

\subsection{Taxonomy of Multimodal Prompting Augmentations}

\revadd{We conducted an additional meeting with the experts to discuss findings from our studies and summarize them into a taxonomy of multimodal prompting augmentations for visualization authoring tasks (detailed in \cref{fig:prompting-taxonomy}).}

\revadd{We observed that visual prompts resolve ambiguity and complement text prompts from three main aspects.
First, visual prompts specify \textbf{spatial constraints}, including {spatial placement} for positioning chart elements and {geometric filtering} for expressing visual regions or boundaries. 
Second, they ground \textbf{local references}, including {element grounding} for binding deictic expressions to concrete chart objects and {in-situ semantic rebinding} for revising a local semantic mapping while preserving the surrounding chart structure. 
Third, they express \textbf{visual design} intent, including {structural composition} for externalizing multi-view layouts and {visual exemplification} for communicating style or encoding preferences through concrete examples. 
Together, these six types characterize how multimodal prompts complement natural language by converting spatial, referential, semantic, structural, and stylistic intent into constraints the model can observe and reason about.}

\revadd{This taxonomy also explains the role of VisPilot's guided interaction design. The system keeps the flexibility of free-form sketching and augments each visual cue with lightweight grounding information, including selected elements, widget types, action descriptions, and the current visualization specification. These signals connect visual input actions to specification-level operations. As a result, multimodal prompts provide a complementary authoring strategy: text prompts communicate abstract intent and precise values, while visual prompts ground local, spatial, structural, stylistic, and semantic constraints in the current visualization context.}

\section{Discussion}
The results of the case studies and user evaluation provide valuable insights into the design of multimodal prompting approaches for visualization authoring tasks.
We summarize the implications as follows.

\subsection{Strengths and Limitations of Prompting Modalities}
Our experiments highlights how each modality contributes to the authoring experience and where their boundaries lie.

\vspace{2pt}
\noindent
\textbf{Visual prompts facilitate LLMs in understanding user intent and improve generation accuracy.}
Our study showed that visual prompts significantly enhance how LLMs interpret user intentions, especially for complex visualization requirements. Sketches that conveyed chart types, element positioning, and visual encoding relationships led to visualizations that more closely matched user intentions compared to text-only prompts. Visual inputs provided direct spatial representations that text often struggled to communicate clearly, reducing ambiguity and eliminating the need for LLMs to make multiple inferences that could lead to misinterpretations. 
This finding demonstrates the potential of visual prompts to improve the accuracy of LLM-generated visualizations, particularly in scenarios where users need to convey intricate visual intent or spatial relationships.

\vspace{2pt}
\noindent
\textbf{Visual inputs reduce the effort for expressing sophisticated visualization demands.}
Our study revealed that visual prompts significantly reduce the cognitive burden of articulating complex visualization requirements. Participants utilized visual prompts more intuitively and efficiently than text prompts when communicating spatial relationships, layout preferences, and design modifications. This was especially notable for tasks requiring precise element positioning such as legends or annotations, where a simple visual indicator achieved what would otherwise necessitate multiple textual exchanges. The multimodal approach allowed users to express intent through the most natural modality for each visualization aspect, reducing the overall effort required.

\vspace{2pt}
\noindent
\textbf{Text prompts excel in parameter specification and conceptual guidance.}
While visual prompts suit spatial relationships, text offers greater efficacy for precise parameter values and high-level analytical goals. Participants exhibited increased efficiency when textually specifying numeric parameters (\eg ``set opacity to 0.7'' or ``use a logarithmic scale for the y-axis'') rather than drawing visual representation of these concepts. Similarly, text proved more effective for communicating abstract visualization objectives like ``show the correlation between variables'' or ``highlight the outliers''. Future systems should integrate text prompts for precise specifications with visual inputs for spatial and design elements, establishing a complementary multimodal framework.

\vspace{2pt}
\noindent
\textbf{Direct manipulation enables efficient iterative refinement.}
Our study demonstrated that direct manipulation of visualization elements significantly enhanced the refinement process. Participants strongly preferred this iterative design paradigm, which enabled them to progressively refine existing visualizations through visual interactions rather than verbal descriptions. This was particularly evident in tasks requiring fine-grained adjustments such as repositioning legends, modifying mark properties, or altering axis scales, as visual actions like drawing arrows for movement or crossing out elements for removal were far more efficient than composing detailed textual instructions.  Therefore, it is beneficial to integrate direct manipulation operations along with multimodal inputs, which can provide seamless transitions between creation and iteration phases with appropriate visual guidance.

\vspace{2pt}
\noindent
\textbf{Lengthy instructions lead to LLM misinterpretation regardless of prompt modality.}
We observed that complex instructions with multiple requirements frequently resulted in partial implementation by the LLM, with certain aspects prioritized while others were overlooked. This phenomenon was pronounced when instructions contained conflicting or ambiguous specifications. For example, detailed visualization requirements with multiple design constraints often led to selective implementation. Similarly, visual prompts with excessive annotations overwhelmed the LLM's processing capacity. Our findings indicate that concise, focused instructions produce superior outcomes, and complex requirements should better be decomposed into sequential interactions rather than consolidated within a single prompt.

\subsection{Opportunities in Multimodal Prompting}
We further explore the broader opportunities that multimodal prompting brings to visualization authoring and implications for future research.

\vspace{2pt}
\noindent
\textbf{Balance between text and visual prompts.}
Our study suggests a complementary division of labor between text and visual prompts. Text is useful for data transformations, parameters, and high-level goals, while visual prompts better capture spatial layouts and local design details. Future systems should therefore support fluid transitions between modalities rather than treating them as separate interaction paradigms or optimizing each modality of prompts~\cite{Feng2024PromptMagician} in isolation.

\vspace{2pt}
\noindent
\textbf{Visual context helps improve LLM reasoning capabilities.}
Integrating richer visual information into context may better support LLMs' reasoning about visualization states, user intent, and specification constraints. In our study, visual prompts supplied spatial and structural cues that were difficult to convey through text alone, but ambiguous sketches or obscure colors were still sometimes misinterpreted. Future work should improve visualization-specific visual reasoning through model training, fine-tuning, or prompt engineering techniques.

\vspace{2pt}
\noindent
\textbf{Multimodal prompt design can inspire brainstorming.}
Some participants used multimodal prompting for exploratory ideation, starting with rough sketches and refining them with visual and textual inputs.
This suggests that visual prompts can support early-stage authoring when users have a partial design concept but have not yet settled on a precise specification.
Future work could further support this process through quick sketching, visual variations, and easy comparison of alternatives, and may inform multimodal prompting in other creative domains such as writing~\cite{Liu2024SPROUT} and image painting~\cite{Hang2025CCA}.

\subsection{Study Limitations}

Our study has several limitations. It primarily used structured datasets and controlled tasks, which may face challenges in unstructured datasets and open-ended real-world workflows. \revadd{The expert evaluation involved only two domain experts, providing qualitative insight into professional authoring patterns rather than statistically generalizable validation. Our formative analysis also relies on a legacy natural language visualization corpus, and future work should examine how users adapt their language and multimodal behaviors in modern LLM-based authoring workflows.} As LLMs evolve to handle larger context windows and complex data reasoning, and integrate advanced data agent techniques~\cite{Weng2025DataLab}, our framework may become more robust and scalable in broader scenarios.

\section{Conclusion}
In this paper, we present VisPilot, a novel approach that addresses the fundamental limitations of text-only prompts for visualization authoring with large language models. Through our empirical study, we identify three key challenges in text-only prompting: limited expression of visual intent, inadequate guidance of LLM behavior, and misaligned human-LLM design preferences. 
Our multimodal prompting framework directly addresses these limitations by enabling users to express their visual intent through complementary visual and textual inputs. 
We develop the VisPilot system and conduct a controlled user study and an expert evaluation to validate its effectiveness.
The results of studies demonstrate that multimodal prompting outperforms text-only prompting in terms of accuracy in task completion and user satisfaction. 
These findings provide strong evidence that incorporating visual prompts can effectively enhance the LLM-based visualization authoring workflow. 
As multimodal large language models continue to evolve, we believe the paradigm of multimodal prompting will become increasingly important for visualization authoring scenarios, making data visualization more accessible while preserving the expressivity and control that visualization authors require.

\section*{Acknowledgments}
\noindent The authors wish to thank reviewers for their valuable feedback.
The authors acknowledge the use of AI tools for language polishing and proofreading of this manuscript.

\bibliographystyle{IEEEtran}
\bibliography{template}

\begin{IEEEbiographynophoto}{Zhen Wen} is a Ph.D. candidate at the State Key Lab of CAD \& CG, College of Computer Science and Technology, Zhejiang University. He received the B.E. degree in Software Engineering from Zhejiang University of Technology in 2021. His research interests include visual analytics and LLMs.
\end{IEEEbiographynophoto}

\begin{IEEEbiographynophoto}{Luoxuan Weng} is a Ph.D. candidate at the State Key Lab of CAD \& CG, College of Computer Science and Technology, Zhejiang University, where he also received his B.E. degree in Software Engineering in 2021. His research interests include visual analytics, human-centered computing, and LLMs.
\end{IEEEbiographynophoto}

\begin{IEEEbiographynophoto}{Yinghao Tang} is a Ph.D. candidate at the State Key Lab of CAD \& CG, College of Computer Science and Technology, Zhejiang University. He received the B.E. degree from Sichuan University in 2023. His research interests include large language models.
\end{IEEEbiographynophoto}

\begin{IEEEbiographynophoto}{Runjin Zhang} received the B.E. degree from the College of Computer Science and Technology, Zhejiang University in 2025.
\end{IEEEbiographynophoto}

\begin{IEEEbiographynophoto}{Yuxin Liu} is an undergraduate student at the College of Computer Science and Technology, Zhejiang University.
\end{IEEEbiographynophoto}

\begin{IEEEbiographynophoto}{Bo Pan} is a Ph.D. candidate at the State Key Lab of CAD \& CG, College of Computer Science and Technology, Zhejiang University, where he also received his B.E. degree in 2022. His research interests include visual analytics and large language models.
\end{IEEEbiographynophoto}

\begin{IEEEbiographynophoto}{Minfeng Zhu} is a tenure-track assistant professor at the School of Software Technology, Zhejiang University. He received his Ph.D. Degree in Computer Science from the State Key Laboratory of CAD\&CG, Zhejiang University in 2020. His research interests include artificial intelligence and visual analysis. For more information, please visit: https://minfengzhu.github.io.
\end{IEEEbiographynophoto}

\begin{IEEEbiographynophoto}{Wei Chen}
is a professor in the State Key Lab of CAD\&CG at Zhejiang University. His current research interests include visualization and visual analytics. He has published more than 80 IEEE/ACM Transactions and IEEE VIS papers. He actively served in many leading conferences and journals, like IEEE PacificVIS steering committee, ChinaVIS steering committee, paper cochairs of IEEE VIS, IEEE PacificVIS, IEEE LDAV and ACM SIGGRAPH Asia VisSym. He is an associate editor of IEEE TVCG, IEEE TBG, ACM TIST, IEEE T-SMC-S, IEEE TIV, IEEE CG\&A, FCS, and JOV. More information can be found at: http://www.cad.zju.edu.cn/home/chenwei.
\end{IEEEbiographynophoto}

\end{document}